\documentclass[useAMS,usenatbib]{mn2e}

\title[Ion-proton model]{Ion-proton model of pulsar radio emission: a summary}
\author[P. B. Jones]{P. B. Jones\thanks{E-mail:
peter.jones@physics.ox.ac.uk}  \\
University of Oxford, Department of Physics, Denys Wilkinson Building,\\
Keble Road, Oxford OX1 3RH, U.K.}

\begin{document}

\date{}

\pagerange{\pageref{firstpage}--\pageref{lastpage}}
\pubyear{}

\maketitle

\label{firstpage}

\begin{abstract}

More than twenty papers on the development of this model have been published in Monthly Notices of the Royal Astronomical Society from 2010 to the present. Whilst some contain work that is essential for the development of the model, others are less so.  This present paper is a summary, citing only the former set of papers and the observational phenomena to which the model relates.

\end{abstract}


\section{Introduction}

Following the discovery of pulsars by Hewish et al (1968) early seminal papers (Gold 1968, Goldreich \& Julian 1969, Radhakrishnan \& Cooke 1969) recognized them as rotating magnetized neutron stars and set out some basic properties of their magnetospheres. The conclusions, a collision-free charge-separated plasma corotating within the light cylinder remain essentially good in that region but externally, the development of the force-free approximation and the prediction of a magnetic-equatorial current sheet by Contopoulos, Kazanas \& Fendt (1999) has modified our view of the magnetosphere as a whole.  Within the light cylinder it
is divided into open and closed sectors of which the open contains all those magnetic flux lines that leave the neutron-star surface and cross the light cylinder. The sign of the corotational charge density $\sigma_{GJ}$ at the polar cap of an approximately dipole field is the negative of the sign of  ${\bf \Omega}\cdot{\bf B}$, in which the neutron-star spin is ${\bf \Omega}$ and ${\bf B}$ is the local magnetic flux density (Goldreich \& Julian 1969).  Particles of these signs, if accelerated, leave the neutron-star surface atmosphere and magnetosphere by crossing the light cylinder and can be lost to the interstellar medium. The sign of $\sigma_{GJ}$ is a little discussed but significant parameter in the description of any specific pulsar, and there appears to be no reason why either particle sign should be preferred at formation. The mechanism of acceleration remained a problem for two decades until Beskin (1990) and Muslimov \& Tsygan (1992) showed that an acceleration field $E_{\parallel}$ exists above the polar cap as a consequence of the Lense-Thirring effect.

Loss of particles of a particular sign from the open sector creates an immediate problem. The neutron star and magnetosphere has to maintain, over time intervals of the order of the rotation period $P$, a stable electrostatic potential with respect to the interstellar medium.  A fraction of the particles from the open sector may re-enter the light cylinder, but it would be unreasonable to expect all particles accelerated to high energies to do so. Thus the radial electric field component in the vicinity of the light cylinder must adjust to a level which allows the escape of particles of the opposite sign. Specifically, the sign of $\sigma_{GJ}$ determines the electric polarity and is a major factor in the structure of the magnetosphere.

High-energy particle currents are the only possible source of pulsar emission in the electromagnetic spectrum; above the polar cap for radio emission and close to the light cylinder in the case of optical, X-ray and gamma-ray emissions.  This division appears good except in the case of the Crab nebular pulsar (Lovelace \& Tyler 1968), {\it sui generis} owing to its age.

The relation between emission frequency band and particle type is simple.  Optical, X-ray and gamma-ray emissions are incoherent and for observable intensities the particles must be leptons.  Radio emission is coherent and requires charge or current density fluctuations on length scales of the order of the wavelength emitted.  Thus the nature of the particles forming the collective fluctuation, leptons or baryons, is not important.

A stable electrostatic potential  implies, for example, a flux of electrons into the region beyond the Y-point if the open-sector out-flow is of positive particles. (The Y-point is the point of contact of the surface separating open and closed sectors with the light cylinder.)  For neutron stars of the opposite polarity this outflow would be of baryons and, in the absence of pairs, there would be no possibility of it being the source of incoherent emission beyond the Y-point. Contopoulos (2016) has argued that the present view of the whole magnetosphere is valid in the absence of electron-positron pairs from any source. But the above considerations are completely vitiated if pair production is present in any part of it.  Very many authors have made this assumption {\it ab initio} based on the possibility that polar-cap flux-line radii of curvature  are small enough to allow single-photon magnetic conversion of curvature radiation. (A reasonable lower limit for this quantity is provided by the condition that a curvature photon produced in the open sector should convert  within that sector.)  Whilst this could be relevant for the younger set of normal pulsars, the assumption cannot be valid, for example, in the case of the rotation period  $P = 8.5$ s pulsar J2144-3933 (Young, Manchester \& Johnston 1999) as noted by the authors themselves.  It also fails in the millisecond pulsars (MSP) for which a test for pair creation exists (Jones 2021) but has not been shown to be satisfied as far as the present author is aware.  These considerations should be viewed in the context that many observers have commented on the broad similarity of pulsar radio emission within the population of both pulsars with periods of the order of $1$ s that are referred to as normal, and the MSP. The extensive literature on pair production in pulsars has been reviewed in two recent papers (Beskin \& Litvinov 2022; Beskin \& Istomin 2022) but the subject is not further discussed here.

In order to reveal the structure of the complete pulsar magnetosphere a rational starting point would be an understanding of one or more emission processes. Coherent radio emission is the most observed and has many characteristic phenomena.  These are summarized in Section 2.  The development of the ion-proton model has been directed to establishing firstly, the physical particle reactions that determine the nature of the plasma accelerated from the polar
cap and then its instability that in coupling with the radiation field results in the radio emission.  More than 20 papers have been published in Monthly Notices of the Royal Astronomical Society in that direction from 2010 to the present.  The purpose of the present paper is to provide a summary of that work.

\section{Pulsar observations}

More than 50 years of observation has produced a catalogue of several thousand pulsars, including normal pulsars and MSP, isolated or binary (Manchester et al 2005). The following Subsections 2.1 - 2.4 attempt to list those phenomena that are common to significant fractions of the population and are likely to reveal the nature of the coherent radio-frequency source that has remained an unsolved problem since 1968.  There are many {\it sui generis} pulsars, in particular the Crab nebula pulsar but the view of this paper is that interesting though they may be, information derived from them does not have the weight of the common phenomena.  The paper also does not favour the view that the radio emission problem is best treated {\it ab initio} exclusively as an exercise in classical electrodynamics.

\subsection{Profiles}

Individual pulse profiles are stochastic in nature so that integrated profiles, obtained by summation over a large number of pulses are the primary recordable source of their properties. Integrations over $10^{3-4}$ s at a given frequency band produce profiles that are stable over time intervals of years.  The spectra have a plateau or maximum in the interval $100 - 400$ MHz and strong negative spectral indices at higher frequencies.  Profile widths (Posselt et al 2021) are weakly frequency-dependent except that some have low-intensity wings at low frequencies.  Integrated pulse shapes can vary with frequency in a complex manner, often having several maxima which, individually, may appear or disappear with changing frequency.  The notch phenomenon is seen occasionally (McLaughlin \& Rankin 2004 ;Dyks, Rudak \& Demorest 2010). This consists of two very sharply defined peaks of intensity that are closely spaced in longitude. Individual pulses vary greatly within a single rotation period and are characterized by modulation indices that, from pulsar to pulsar, can vary by several orders of magnitude becoming large with age.

It is more physically revealing to consider the radio-frequency energy per primary unit charge accelerated in the open magnetosphere in place of the luminosity or flux density.  This varies over several orders of magnitude and can be surprisingly large, typically centred on the $0.1 - 1.0$ GeV interval (see Jones 2014).

\subsection{Polarization}

The emission is polarized: Radhakrishnan \& Cooke (1969) observed that the linear polarization plane rotated through an angle approaching $\pi$ radians as a function of observation longitude, indicating that the radiation electric field is radially polarized above the polar cap.  Recent very extensive surveys have confirmed that this can be regarded as a standard feature of pulsar emission (Oswald et al 2023; Posselt et al 2023; Johnston et al 2023) although often one or more $\pi/2$ discontinuities in polarization position angle are superimposed, each accompanied by zeros in polarization. Circular polarization is also observable in many pulsars, both normal and MSP. Karastergiou et al (2003) observed that the Stokes-V parameter can change sign from pulse-to-pulse.  Karastergiou \& Johnston (2006) observed Stokes-V at 1.4 and 3 GHz and found in general, only modest frequency-dependence (see also Kramer \& Johnston 2008). Dai et al (2015) have published Stokes-V for a set of 24 MSP.

\subsection{Nulls}

The phenomenon of nulls is seen in a significant minority of pulsars. A null refers to the abrupt and usually complete disappearance of radio emission at all observable frequencies followed by its reappearance after an interval of one or more rotation periods.  We refer to Kramer et al (2006) and to Redman \& Rankin (2009) concerning non-stochastic nulling.The distribution of null lengths is a sharply decreasing function of this time interval.

A smaller minority of pulsars have mode changes.  Here, the integrated profile changes shape abruptly, possibly contemporaneously with X-ray emission if that is present in the pulsar. A low level of radio emission is observed in a small minority of pulsar null states indicating, perhaps, that nulls and mode changes are not unrelated. The existence of nulls indicates that quantum electrodynamics is the source of radio emission even if presently unknown particles of cosmological origin are involved. 

\subsection{Subpulses and drift}

The polar cap is usually assumed to be circular and of radius, estimated for a dipole field extending to the light cylinder radius $R_{LC}$ as,
\begin{eqnarray}
u_{0} = \left(\frac{2\pi R^{3}}{cPf(1)}\right)^{1/2}
\end{eqnarray}
(Harding \& Muslimov 2002) where $R$ is the neutron-star radius and the constant $f(1) = 1.368$ for a neutron star of mass $1.4M_{\odot}$. The true shape of the polar cap could be quite different and such variations may be connected with the long-term stable but complex integrated intensity profiles referred to in Section 2.1. 

There are clear indications that subpulses exist in which, at any instant, radio-emitting plasma sources are present above only a fraction of the polar-cap area. The manner in which these drift over the polar cap within time intervals of the order of several rotation periods suggested an ${\bf E}\times{\bf B}$ velocity (see Ruderman \& Sutherland 1975) owing to the presence of a radial electrostatic field above the polar cap satisfying the assumed boundary condition separating open and closed sectors of the magnetosphere.  But more recently, pulsars having simultaneously subpulses drifting in opposite directions have been observed (see Champion et al 2005; Weltevrede, Edwards \& Stappers 2006). Complex behaviour is frequent and explanation remains obscure.

\section{Ion-proton model}

Coherent radio emission certainly requires particle acceleration as an energy source and observations are consistent with this taking place in the open sector of the magnetosphere at a relatively low altitude compared with the light cylinder radius in normal pulsars. Such low altitudes are all that are available in the MSP. The current density in the open sector is a source of disagreement between authors. The Goldreich-Julian current density $\sigma_{GJ}$ in a perfectly corotating magnetosphere is a function of flux-line curvature with respect to the magnetic axis but at low altitudes this can be neglected in an approximately dipole field.  However, some authors, following one-dimensional studies of electron acceleration immediately above the polar cap (Mestel et al 1985; Beloborodov 2008; Bai \& Spitkovsky 2010) assume that the open-sector current density is not necessarily $j =c \sigma_{GJ}$ but is fixed at a value $|j| > |c\sigma_{GJ}|$ by the properties of the whole magnetosphere. At values $|j| < |c\sigma_{GJ}|$ there is no useful acceleration in one dimension.  But at current densities above this limit useful acceleration is technically possible given a system of intermittent electron-positron pair creation (Timokhin \& Arons 2013: Philippov, Timokhin \& Spitkovsky 2020). The model is strictly for $\sigma_{GJ} < 0$ but it is also claimed that the model can be extended to the case of $\sigma_{GJ} > 0$. Pair creation is a condition of the model so that in the latter case no positively charged particle is accelerated outward from the polar-cap surface. The status of this model is not clear at present. A test of its validity in the MSP has been proposed recently (Jones 2021) but to date, there is no published evidence that it is been satisfied. Specifically, Figure 10 in the Third Fermi Catalogue of Gamma-ray Pulsars (Smith et al 2023) shows the phase separation between the radio pulse and the first of a pair of main gamma-ray peaks for both young normal pulsars and MSP.  There is clearly no indication of any grouping in the vicinity of zero  phase separation.

There is also a problem concerning the plasma frequency $\nu_{p}$ in high-multiplicity pair plasmas. It is convenient to measure altitude $z$ in terms a dimensionless parameter $\eta$, $z = (\eta - 1)R$.  In the observer frame it is of the order of
\begin{eqnarray}
\nu_{p} = \gamma \left(\frac{2\alpha B{\rm e}}{\pi mcP\gamma\eta^{3}}\right)^{1/2}     \nonumber  \\
        =3.4(\alpha\gamma)^{1/2}\eta^{-3/2}(B_{12}P^{-1})^{1/2} {\rm GHz}
\end{eqnarray}
in which $\alpha$ and $\gamma$ are here, respectively, the pair-cloud multiplicity and its rest frame Lorentz factor, and $B$ is the polar-cap magnetic flux density at the surface. Magnetic flux density declines as a function of altitude $\eta$ but for commonly assumed values $(\alpha = 10^{3 -5}, \gamma = 10)$, the frequency is high compared with those of maxima in the radio-frequency spectra generally observed which are of the order of $100 - 400$ MHz. We refer also to Melrose \& Gedalin (1999) on this subject.

These two considerations and the fact that coherent radio emission does not require a leptonic source served to motivate the development of the ion-proton model in which $\sigma_{GJ} > 0$. It assumes that asymptotically, $\sigma(\eta) \rightarrow \sigma_{GJ}(\eta)$ and that at lower altitudes, acceleration is produced by the Lense-Thirring effect (Beskin 1990; Muslimov \& Tsygan 1992; see also Harding \& Muslimov 2001, 2002). The total potential difference is of the order of $10^{3}B_{12}P^{-2}$ GeV making pair creation impossible without special assumptions about flux-line curvature in a considerable fraction of normal pulsars. Development of the model has proceeded in two stages: firstly an investigation of the physical processes that determine the nature of particle fluxes at the polar cap and secondly, the ensuing plasma instability that is the source of the radiation.

\subsection{Particle beam formation and acceleration}

The nuclear composition of the neutron-star surface is the most serious unknown.
The most usual assumption is the canonical $Z_{0} = 26$ except that $Z_{0} = 6$ is a possibility following the work on the X-ray spectrum of the Cas A neutron star by Ho \& Heinke (2009). We also assume it consists of normal matter and could be either liquid or solid depending on the particular conditions that are relevant. Surface densities have been calculated by many authors but we mention in particular the work of Medin \& Lai (2006) who found densities of the order of $10^{2-3}$ g cm$^{-3}$ and confirmed that the ion work function is unlikely to interfere with the free movement of ions from the stellar surface into its atmosphere. There is a reasonable case, in general, for assuming that the structure of the atmosphere immediately above the neutron-star surface is determined by its ion and electron components in local thermodynamic equilibrium with the gravitational field and a small internal electric field parallel with the magnetic flux. Its scale height is of the order of $0.1$ cm and its density is dependent on both temperature and ion work function. The important parameter is the ionic charge-to-mass ratio.  Thus the numerically small component of protons cannot be in equilibrium in the atmosphere and hence have a small outward drift velocity (Jones 2013).  On leaving the atmosphere protons are subject to the Lense-Thirring acceleration which is represented (see Muslimov \& Harding 1997) as a modified corotational charge density that is smaller than than the flat-space Goldreich-Julian charge density which is reached asymptotically. Following Muslimov \& Harding,
\begin{eqnarray}
\sigma_{LT}(\eta) = (1 - \kappa(\eta))\sigma_{GJ}(\eta)
\end{eqnarray}
with $\kappa(1) \approx 0.15$, for a typical $1.4 M_{\odot}$ neutron star and tending to zero as $\eta$ becomes large.

Similarly, ions drawn from the atmosphere are subject to acceleration but blackbody photons from the neutron-star surface that is visible to the ion have large photoelectric cross-sections when acceleration causes their energies in the ion rest frame to reach the thresholds for occupied states.  The electrons are reverse-accelerated to the neutron-star surface. This amounts to a simple screening process that controls the actual acceleration field and limits the ion Lorentz factors to small values.  A two-component plasma of ions and protons is similarly limited (in general, the protons are completely ionized in the LTE atmosphere).

The source of the protons is principally the terminal stage of the electromagnetic showers that the reverse-accelerated electrons create in the atmosphere and, possibly, extending into the solid or liquid phase of the surface.  The nature of these showers does not differ greatly in most normal pulsars and the MSP from the $B = 0$ case (see Jones 2010).  The shower photon track length per unit interval of photon energy is known in the neighbourhood of the nuclear giant dipole state as are the cross-sections for its formation and the decay branching ratios for small-to-medium atomic number nuclei.  The proton production rate can be summarized as approximately $ W_{p} = 0.2$ GeV$^{-1}$ of shower energy for wide intervals of $B$ and of ion atomic number $Z_{0}$ (Jones 2010) and does not differ much from the $B = 0$ values.
Proton production has a maximum at a shower depth of about $10$ radiation lengths and the diffusion time to the upper part of the LTE atmosphere is of the order of $1$ s (Jones 2013).

The blackbody radiation producing photoelectrons is distinct from that of the polar cap and is that of the whole neutron-star surface visible to the ion. Whole-surface temperatures as low as $2\times 10^{5}$ K are effective. Radiation from the polar cap though generally of higher temperature is less effective because it subtends small angles to the ion velocities at which the Lorentz transformations to the ion rest-frame are unfavourable.

The changes in a nucleus, initially of atomic number $Z_{0}$ as it moves within the shower interval of depth before it reaches the neutron-star surface are as follows.  Proton creation reduces $Z_{0}$ to $Z$ at the time it reaches the part of the LTE atmosphere from which its acceleration becomes possible.  The important parameter is then $\tilde{Z}_{s}$, the ion charge, which is a function of polar-cap temperature.  After losing photoelectrons, this becomes $\tilde{Z}_{\infty}$ and $\tilde{Z}_{s} = (1 - \kappa/2)\tilde{Z}_{\infty}$, in which $\kappa$ is the Lense-Thirring factor.  Here, $\kappa \approx \kappa(1)$. The number of protons created per ion is then,
\begin{eqnarray}
Z_{0} - Z = \kappa \epsilon W_{p}\tilde{Z}_{\infty}/2,
\end{eqnarray}
in which $\epsilon$ is the mean energy reached by a reverse electron, and all the parameters $Z, \tilde{Z}_{s}$ and $\tilde{Z}_{\infty}$ are averages in this time-independent expression.  The charge density is then  approximately equal to $\sigma_{LT}$.  Screening of the acceleration field is almost complete so that ions have quite modest Lorentz factors during growth of the Langmuir mode.

\subsection{The Langmuir mode}

Physical conditions are not uniform over the polar cap owing to the boundary condition on the surface separating the open from the closed sectors of the magnetosphere, and possibly to the polar cap having an irregular shape.  Consequently, if coherence in the emission process is sought it may be preferable to consider coherence within a narrow bundle of flux lines rather than coherence over the cross-section of the whole complete open sector.  Thus an unstable Langmuir mode is appropriate. Asseo, Pelletier \& 
Sol (1990) investigated a quasi-longitudinal Langmuir mode in an electron-positron plasma, followed (Asseo \& Porzio 2006) by the formation of Langmuir solitons as a source of radio emission. However, the problems arising with a plasma of continuous-energy particles are obviated completely in a two-component baryonic plasma with discrete velocities for each component. Its dispersion relation is,
\begin{eqnarray}
1 - \frac{\omega^{2}_{s1}}{(\omega - qv_{1})^{2}}
 - \frac{\omega^{2}_{s2}}{(\omega - qv_{2})^{2}} = 0
\end{eqnarray}
for a two-component baryonic plasma in which,
\begin{eqnarray}
\omega^{2}_{i} = \frac{4\pi n_{i}Z^{2}_{i}{\rm e}^{2}}{m_{i}\gamma^{3}_{i}}
\end{eqnarray}
with $i = s1,s2$.  The velocities, Lorentz factors, number densities, and particle masses of each component are $v_{i}$, $\gamma_{i}$, $n_{i}$ and $m_{i}$
respectively.  The protons, and approximately the ions, have each been accelerated through the same potential difference and hence the velocities are discrete and different.  The mode angular frequency and wave number are $\omega$ and $q$ respectively, and these, with all parameters in equation(5), have observer-frame values.  Equation (5) does not permit Landau damping which would be present in a continuum of particle velocities.  It has two real roots and a conjugate pair. The unstable mode growth rate is adequate for the attainment of non-linearity in pulsars (see Jones 2012).  The energy source relies on the fact that any longitudinal section of the beam, not under acceleration and considered as an isolated system, has constant momentum but is not in its lowest energy state which would be one in which all particle velocities are identical.  Therefore, coupling of the mode with the radiation field moves the disturbed beam velocities towards each other whilst any acceleration field tends to separate them and thereby support the rate of radio-frequency energy emission.

The mode requires two particle beams of different charge-to-mass ratio. The proton flux at the neutron-star surface is a definite consequence of the reverse-electron flux.  However, protons from other sources, howsoever produced, may be present and required if the neutron-star surface temperatures are low. A further possibility, relevant particularly in the case of small atomic numbers $Z_{0}$, is that ions of identical nuclear charge but quite different charge-to-mass ratio can be present in the atmosphere.

Modelling (Jones 2020a) confirms that the local ion-proton fluxes are stochastic in nature and this introduces subpulse formation: an area of the polar cap in which the growth rate of the mode is particularly large at some instant of time.  Its linear dimension must be at least a small multiple of the Langmuir mode wavelength: whereas equation(5) assumes an infinite plasma.  This indicates that subpulses should be perhaps no more than an order of magnitude smaller in area than the polar cap. We refer to this later in Section 4.

The presence of an outgoing positron flux would require extension of equation (5) by a further term of identical structure.  But the positron Lorentz factor, after passage through the accelerating potential, would be so large as to render the term negligible compared with unity.  Reverse electrons similarly, and with a change of sign in velocity, have negligible effect.

Mode growth rates are large at small Lorentz factors (see Jones 2012, 2022) and there are no obvious problems which could prevent the attainment of non-linearity.  Owing to this and to the dependence of the frequency interval of any individual mode on the parameters in equation (5), which may vary according to the state of the particular area of the polar cap supporting it, individual pulses and certainly the integrated pulse profiles should appear as continua.

The emission mechanism described (Jones 2023) is concerned principally with coherence parallel with polar-cap flux lines and because baryonic plasmas have dielectric tensor components very close to zero or unity, the mode emission can be treated to a first approximation as though in vacuo. This point is also significant in the treatment of circular polarization.  The mechanism described here may be the first in which each step in the process of emission has a clear physics basis.

\section{The Langmuir mode and observation}

The observations summarized in Section 2 are the result of half a century of observational work.  Those in Sections 2.1 and 2.2 were established quickly and are observed in a substantial fractions of the population. In consequence, the ion-proton model has been developed {\it ex post facto} so that prediction of any new phenomenon not yet observed is difficult.

\subsection{Profiles and Nulls}

The shortest time-scale expected from the longitudinal Langmuir mode is of the order of $10^{-5}$ s, determined by the interval of altitude within which the mode radiates strongly.  The longest is likely to be associated with the relaxation time $\tau$ for proton diffusion from creation to the top of the LTE atmosphere.  The presence of subpulses is easily accommodated.

The fact that the polar cap in pulsars of medium age is in a stochastic state indicates that at any instant, growth of a Langmuir mode is likely to be restricted to certain regions.  Elsewhere there can be too many protons in the upper part of the atmosphere to establish conditions for mode growth.  These regions of growth appear as subpulses and the time-scale for their existence and possible movement is the proton diffusion time $\tau$.

Integrated profiles are stable over time intervals of several years and frequently have curious asymmetric shapes.  This is beyond the scope of the ion-proton model but may indicate that the common assumption of a circular polar cap is incorrect in many cases: the shape is more likely to be determined by the presence locally of high-order multipole field components.  Also there appears to be no reason why the polar cap should not consist of more than one disconnected area.

Radiation from the Langmuir mode has a narrow angular distribution and, for axial symmetry, zero intensity parallel with the mode axis (see Jones 2023).  Connection with notches in the integrated profile is an obvious possibility when the observer's arc of transit passes near to the mode axis. The sharp nature of the notches as functions of observer longitude is consistent with coherence principally parallel with ${\bf B}$.

Observation of nulls is expected because the emission of radiation depends on mode formation which requires two beams of different charge-to-mass ratio.  Excessive production of protons, over the polar cap, or some part of it, willresult in no growth.  Proton loss from an element of polar cap occurs at a rate limited, though not exactly, by the Goldreich-Julian current density and any excess of protons forms a layer at the top of the LTE atmosphere.  Only when this is exhausted can mode growth re-commence.

The model polar cap (Jones 2020a) demonstrates a number of interesting features: the variations of the phase-resolved intensity modulation index (as defined by Jenet \& Gil 2003), the null fraction and the distribution of null lengths, as functions of surface temperature $T_{s}$ and period $P$ for various values of $B$ and $Z_{s}$.
In the chaotic state of the polar cap, the partially screened acceleration potential is a time dependent function of position but the assumed boundary condition between open and closed sectors broadly remains so its time-averaged values are larger at the polar-cap centre than near the boundary at $u = u_{0}$.  The mode amplitude growth exponent is $\propto \gamma^{-3/2}_{A,Z}$ or $\gamma^{-3/2}_{p}$ so that increased Lorentz factors can reduce emission intensity to negligible values.  This is consistent with the presence of conal and core-profiles.  The higher Lorentz factors in the core region reduce mode growth rates in that region, leaving only the conal profile which is frequently seen in older pulsars.

The Lorentz factor at which blackbody photons, of temperature $T_{s}$, reach a photoelectric threshold in the ion rest frame is $\gamma_{c} \propto T_{s}^{-1}$.  Thus at constant $B$ and $P$ there is a critical $T_{s}$ below which the mode amplitude as a source of emission is negligible.  As $T_{s}$ approaches its critical value, the null fraction, mean null length and intensity modulation index increase until the null fraction approaches unity.  At constant $B$ and $T_{s}$, the null fraction is not a rapidly varying function of $P$, but mean length and intensity modulation indices decrease quite rapidly as $P$ increases.

\subsection{Polarization}

Radhakrishan \& Cooke (1969) observed the rotation of the linear polarization plane through an angle approaching $\pi$ radians as a function of longitude.  This is now a well-established phenomenon (Oswald et al 2023, Posselt et al 2023) and is most obvious in young pulsars that have high linear polarization.
Circular polarization is also present in profiles, usually at a much lower intensity. It is widely assumed to be a consequence of the magnetosphere through which the radiation passes to the observer. A complex limiting polarization argument can be made in the case of an electron-positron magnetosphere (Melrose \& Stoneham 1977; Cheng \& Ruderman 1979; Barnard 1986; Lyubarskii \& Petrova 1999; Petrova \& Lyubarskii 2000; Beskin \& Philippov 2012 and other authors who drew on the work of Budden 1952) but it does not predict the change in sign of the Stokes V-parameter that is frequently observed within a profile. Also it fails to describe the V-parameter observed in the MSP.

An open-sector magnetosphere of baryonic particles much simplifies the problem of circular polarization (Jones 2016).  Components of the dielectric tensor are extremely close to zero or unity and refractive indices are such that physical separation in the propagation of O and E-modes is negligible.  In the neutron-star rest frame ${\bf B}$ is the field, here assumed dipole, and ${\bf k}$ is the radiation wave-vector that traverses the polar cap during observation  Then in the evaluation of ${\bf k}\times{\bf B}$, the unit vector components $\hat{k}_{x,y}$ and $\hat{B}_{x,y}$ are very small compared with unity.  To the first order of smallness, the direction of the E-mode electric vector is given by
\begin{eqnarray}
tan\theta_{E} = \frac{\hat{B}_{x} - \hat{k}_{x}}{\hat{k}_{y} - \hat{B}_{y}}
\end{eqnarray}
and for the O-mode, $tan\theta_{O} = -cot\theta_{E}$.  Here the axes can be chosen so that $\hat{k}_{y}$ is almost exactly a constant during traverse but $\hat{k}_{x}$ changes sign on passage past the magnetic axis (Jones 2016).  In a later paper (Jones 2020b), the Stokes parameters I, L, and V have been evaluated for the particular case of the
$8.5$ s pulsar J2144-3933 using the above expressions, and for more general cases.  Although the assumed mechanism for the source of radiation assumed in that paper does differ from that which is now believed to be correct (Jones 2023) these evaluations still stand. The general conclusion is that the observed circular polarization is well described by the ion-proton model (Jones 2020b) as is the dependence of the sign of V on the gradient of the linear polarization position-angle sweep (Johnston \& Kramer 2019, Desvignes et al 2019).

Linear polarization has, in many pulsars, further complications.  The polarization position angle as a function of longitude can have one or more $\pi/2$ discontinuities each accompanied by a zero in Stokes-L.  This has been described by many authors as a transition between two orthogonal modes above the polar cap, the nature of the modes being unspecified.  The most recent work phenomenological is that of Oswald et al (2023) in which the modes are assumed to be partially coherent.

However the explanation for these observations is immediate once it is recognized that at wave-vector ${\bf k}$, an observer at any instant receives radiation from an area of polar cap that is finite and of area only an order of magnitude smaller than the polar cap (Jones 2017a).  Thus it is possible to divide the polar cap, in principle, into two non-unique sources of polarized intensity $I_{1}p_{1}$ and $I_{2}p_{2}$.  Again in principle, it would be possible to define a curve on which an observer would see that $I_{1}p_{1} = I_{2}p_{2}$. The same observer, looking at polarization position angles $\phi_{1,2}$, would then see that at some point on the curve between the exterior of the polar cap and proximity to the magnetic axis, the condition $\phi_{1} - \phi_{2} = \pi/2$ would be satisfied.  The division of the polar cap into two sources can be modified by an incremental change and the above argument repeated, so establishing the existence of one or two curves on which the $p = 0$ condition exists and that may, but not universally, be intersected by an observer's arc of transit.

Oswald, Karastergiou \& Johnston (2023) have described polarization in terms of a superposition of two orthogonal modes that are partially coherent, an interesting and novel approach to the problem.  The ion-proton model description of pulsar polarization is quite consistent with this.  The previous paragraph describes the $\pi/2$ jump phenomenon in different but equivalent terms. The ion-proton model is also consistent with partial coherence.  Circular polarization generated by a single subpulse or area of of a single Langmuir mode is certainly a consequence of the natural coherence of emission from that mode.  But more general coherence over the whole area of a polar cap is difficult to accept other than the unlikely circumstance of the mode occupying the whole area.

\subsection{Subpulse drift and mode changes}

The existence of subpulses in the ion-proton model described in Section 3 is clear but the description of subpulse drift has proved difficult. Drift speeds appear not inconsistent with an azimuthal ${\bf E}\times{\bf B}$ velocity above the polar cap, but the presence of opposite drift velocities in the same pulsar casts doubt on this explanation.  A characteristic time in the existence of a subpulse is the mean time for proton diffusion from its point of creation to the top of the atmosphere.  This is not well-known but could be of the order of one second.  Consequently, it is possible for a subpulse to move towards an adjacent area in which the ion-proton ratio is more favourable for Langmuir mode growth.  In this way, it is possible that the same path, possibly starting near the polar-cap boundary $u = u_{0}$, may be repeated.  But modelling of such behaviour is difficult.

Similarly, it has not been possible to model mode changes in any satisfactory way.  For a magnetic axis subtending some finite angle with ${\bf \Omega}$ the condition of the polar cap can be complex with the possibility of pair creation by inverse Compton scattering (see Hibschman \& Arons 2001) in a section of it. The more stable mode would presumably be that in which the acceleration field is more completely screened.  But it has not so far proved possible to establish satisfactorily the conditions for stable co-existence of the two modes.  However, there is some confidence that the pre-conditions exist in the ion-proton model.

\section{Conclusions}

The ion-proton model has two merits: it has a basis in physical processes that are well understood and does not rely on neutron-star parameters having values within very limited intervals.  The polar-cap magnetic flux density is not critical and the model is satisfactory for MSP as it is for normal pulsars with fields of the order of $10^{12}$ Gauss.  The photoelectric process is self-regulating and consistent with the principle that screening should occur to the maximum extent possible. Fortunately,the process is not sensitive to photoelectric cross-sections which are not well known: a change in cross-section merely results in a small compensating change in ion Lorentz factor.

It is well-known that the longitudinal Langmuir mode in an infinite plasma does not couple with the radiation field. But the mode confined to a column of limited transverse dimensions is not so constrained, and its radiation is principally longitudinally coherent as is required, for example, by the observation of the notch phenomenon.

The objective has been to start by describing particle acceleration in the ${\bf \Omega}\cdot{\bf B} < 0$ case and to consider first phenomena that are seen commonly in the majority of pulsars. Pulsars that are individually {\it sui generis} are often less revealing, except, perhaps J2144-3933.  An example is the Crab nebula pulsar.  Its principal radio and gamma-ray emissions coincide in longitude which suggest that this radio source is  near the light cylinder.  Its sign of ${\bf \Omega}\cdot{\bf B}$ is unknown: the only indication of polar-cap emission consistent with the ion-proton model is a relatively weak low-frequency profile about $40$ degrees prior to the main pulse which also has circular polarization.

Neutron stars with ${\bf \Omega}\cdot{\bf B} >0$ must presumably exist but it is not obvious that they populate the bulk of the $P - {\dot P}$ diagram.  The time-scales for magnetospheric evolution could be of the order of $1$ Myr and it is quite possible that pulsars could appear, disappear, or re-appear on this diagram without being observed.  The emission of gamma-rays is a not unrelated question.  The sign of ${\bf \Omega}\cdot{\bf B}$ is of general interest because it does affect the structure of the whole magnetosphere including regions beyond the light cylinder.  Thus in the negative case, there must also be a loss of electrons from beyond the Y-point. In this process, the electron flux is large enough to be the source of gamma-ray emission (see Jones 2017b).  In the positive case, positive particles must be lost beyond the Y-point and, in the absence of pair creation, only baryons are available. Consequently, pair creation must occur in some part of the magnetosphere to make gamma-ray emission possible at Fermi-LAT energies.

\bsp

\label{lastpage}

\end{document}